# Linear conduction in N-type organic field effect transistors with nanometric channel lengths and graphene as electrodes

F. Chianese[1,2,a)], A. Candini[3,4, a)], M. Affronte[3,5], N. Mishra[6], C. Coletti[6], A.Cassinese[1,2]

[1] *Physics Department, University of Naples 'Federico II', Piazzale Tecchio, 80, I-80125 Naples, Italy.*

[2] *SPIN-CNR Division of Naples, Piazzale Tecchio, 80, I-80125 Naples, Italy.*

[3] *Centro S3, Istituto Nanoscienze - CNR, via G. Campi 213/A , 41125 Modena. Italy*

[4] *Istituto per la Sintesi Organica e la Fotoreattività (ISOF) – CNR, Via P. Gobetti, 101 - 40129 Bologna, Italy*

[5] *Dipartimento di Scienze Fisiche Informatiche e Matematiche, Università di Modena e Reggio Emilia, via G.Campi 21$^1$3/A, 41125, Modena, Italy.*

[6] *Center for Nanotechnology Innovation @ NEST, Istituto Italiano di Tecnologia, Piazza San Silvestro 12, 56127 Pisa, Italy*

In this work we test graphene electrodes in nano-metric channel n-type Organic Field Effect Transistors (OFETs) based on thermally evaporated thin films of perylene-3,4,9,10-tetracarboxylic acid diimide derivative (PDIF-CN2). By a thorough comparison with short channel transistors made with reference gold electrodes, we found that the output characteristics of the graphene-based devices respond linearly to the applied biases, in contrast with the supra-linear trend of gold-based transistors. Moreover, short channel effects are considerably suppressed in graphene electrodes devices. More specifically, current on/off ratios independent of the channel length ($L$) and enhanced response for high longitudinal biases are demonstrated for $L$ down to ~140 nm. These results are rationalized taking into account the morphological and electronic characteristics of graphene, showing that the use of graphene electrodes may help to overcome the problem of Space Charge Limited Current (SCLC) in short channel OFETs.

Organic electronics have gathered a great deal of attention due to the possibility to fabricate flexible and low cost devices for several applications.[1–4] Despite considerable advances in the understanding and controlling the fundamental mechanisms, several bottlenecks still restrict the miniaturization of Organic Field Effect Transistors (OFETs) towards the nanometric scale. In particular, in short-channel architectures the transport mechanism is often dominated by non-linear Space Charge Limited Current (SCLC)[5–8] in which the current is mostly given by charge carriers directly injected from the contacts to the bulk of the organic thin film. This spurious effect reduces

---

the overall performances of the devices at the nano-metric scale, despite the fact that the maximization of the output currents, switching frequencies and minimum supply voltages are directly related to the reduction of the active channel length $L$ [9,10]. Short-channel effects are primarily due to the highly intense electric fields building up at the channel. They hinder charge accumulation at the dielectric interface and manifest in the form of parabolic current-voltage characteristic for channel lengths already at the micrometric scale [11] and high off-state currents due to drain-induced barrier lowering, commonly encountered in inorganic devices as well [12,13]. In this scenario, contact effects at the organic semiconductor/electrode junctions play a key role. Charge carriers injection and extraction mechanisms in OFETs are indeed dictated by the interplay between the net alignment of the Fermi level of the metallic electrodes with the LUMO (HOMO) level of the organic semiconductor, as well as the energetic and morphological order of the organic thin film nearby the contacts[14–16]. Careful choice of materials or chemical treatment of contact electrodes is therefore needed in order to optimize the OFETs response [17], especially in nanometric architectures. Graphene and graphene-based materials have been recently considered as a novel electrode material in field-effect devices based on conventional organic semiconductors[18–22], graphene nano-ribbon[23] and single molecule junctions,[24] taking advantages from its work function tunability, [25] its permeability to transversal electric field and its overall chemical stability [26,27].

In this work, we report on the fabrication and characterization of short channel n-type OFETs with graphene electrodes and with channel length $L$ ranging from 140 nm to 1 µm, with a fixed channel width ($W$=2µm). For direct comparison, devices with an analogous architecture have been fabricated using standard gold contacts in the same $L$ interval. As the organic semiconductor, we used a perylene-3,4,9,10-tetracarboxylic acid diimide derivative, also known as PDIF-CN$_2$ (Polyera ActiveInk$^{TM}$ N1100), which has gained great attention in the last years due to its enhanced air stability and excellent n-type transporting properties[28–30], yielding field-effect mobility values largely exceeding 1 cm$^2$V$^{-1}$s$^{-1}$ in state of the art single crystal devices [31].

As the electrode material, we employed polycrystalline monolayer CVD graphene grown and transferred with a wet-approach, as reported in [32], on a 300 nm-thick SiO$_2$/doped-Si substrate. Graphene sheets under analysis are found to be heavily p-type doped, as it can be seen from the transport characteristics of micrometric strips ($W$=2µm, $L$=4µm) used as a benchmark (Figure S1 in Supplementary Material). Graphene doping was further investigated via Scanning Kelvin Probe Microscopy (Park-XE-100) in ambient condition. Using a grounded gold pad as the reference for the gold-coated AFM cantilever, an average work function of about 4.9 eV for the graphene was determined (Figure S2). Unintentional doping effects in CVD-graphene are commonly encountered and are usually related to substrate surface treatments prior to the graphene transfer [33] or to contamination due to environmental exposure [34]. Devices have been realized starting from the large area graphene sheet by means of Electron-Beam Lithography (EBL) and oxygen plasma Reactive Ion Etching (RIE), employing the underlying doped-Si as the gate in the bottom contact-bottom gate (BC-BG) architecture. A micrometric pattern of gold tracks and probe pads (3/30 nm and 5/50 nm of thermally evaporated Cr/Au, respectively) is over imposed on the nano-patterned graphene, obtaining the structure depicted in Figure 1 (a), (b). An analogous fabrication procedure has been used for the devices with gold electrodes. After the lithographic processes, the samples have been cleaned via N-methylpyrrolidone-solvent bath at 65°C for 4.5 hours, then left in acetone for 2 hours at room temperature and rinsed in isopropanol. This was done in order to eliminate resist contaminants and improve the organic thin film morphology especially on the graphene electrodes. Lastly, the $SiO_2$ substrate has been functionalized by a self-assembled monolayer (SAM) of hexamethyldisilazane (HMDS) by vapor priming (1 hour at 160°C) enhancing in such a way the growth of the organic thin film and the overall performances of the devices.[29]

PDIF-CN2 has been deposited via Organic Molecular Beam Deposition (OMBD) on the substrates kept at 100°C with a deposition rate of about 0.3 nm/min in high vacuum conditions, resulting in a 25 nm thick polycrystalline thin film (Figure1 (c) and (d)). The electrical characterization has been performed in a Janis probe station, in vacuum and dark conditions, employing a Keithley-2612A SourceMeter.

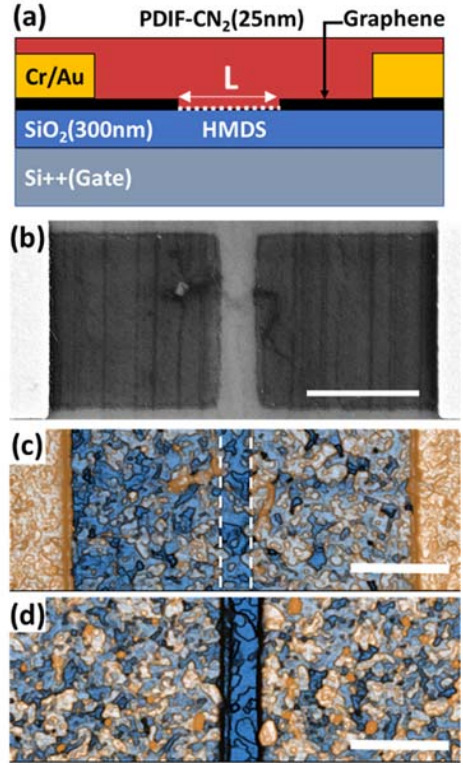

**Figure 1 (a)** Sketch depicting the bottom contact-bottom gate architecture used for the n-type short channel OFETs under analysis **(b)** SEM image of a short-channel FETs based on graphene electrodes ($L=400\pm30$nm). Channel width ($W$) is fixed at 2 μm for every channel length $L$. Enhanced-color AFM topographies of graphene-based transistor (c) and gold-based device (d) with $L=400\pm30$nm. Dashed lines in (c) highlight the graphene/channel interfaces. From AFM analysis a root-mean-square roughness (σ) of ≈1.4nm and ≈1.8 nm are estimated in the channel and on the graphene electrodes respectively, while σ≈1.2nm and ≈1.6 nm are observed in the channel and on the electrodes in gold-based transistors. The scales in (a), (b) and (c), indicated by the white bars, are of 1μm.

Output curves for three out of six different channel lenghts under analysis are reported in Figure 2 (refer to Figures S3 and S4 for the whole set of $L$). A comparative analysis shows substantial differences between the gold-based and graphene-based nano-devices. Focusing on the graphene electrodes (Figure 2(a) to (c)), the drain-source current ($I_{ds}$) responds linearly to the drain-source bias ($V_{ds}$) in the range 0 V<$V_{ds}$<20 V. An obvious current modulation is observed in the case of graphene electrodes for increasing the gate-source bias ($V_{gs}$), indipendently on the channel length $L$. Increasing $L$ from 140±30nm to 1000±30nm, the maximum $I_{ds}$ values are observed to decrease accordingly. Conversely, in the case of gold electrodes (Figure2 (d) to (f)) a supra-linear behavior is evident in the same $V_{ds}$ interval, while the $V_{gs}$ current-modulation deteriorates for $L$ approaching the minimum value of 140 nm (Figure 2 (d)). It should be pointed out that in both architectures current saturation is not achieved for any $L$ due to the relatively thick gate dielectric (300 nm) and its effective charge accumulation. Moreover, despite PDIF-CN2 organic thin films show similar morphologies inside the active channel of graphene and gold-based transistors (Figure 1 (c) and (d)), maximum $I_{ds}$ values was observed to differ of nearly one order of magnitude indicating a major contribution of the contact resistance for the case of graphene electrodes. Contact effects are indeed known to be not uniquely dictated by the energetic alignment at the electrode/organic Schottky interface, but morphological as well as geometrical contributions must be considered too. In fact, the correlation between the molecular crystal domain orientation on the electrodes and the resulting work function and injection barrier has been studied for different configurations, in particular pentacene on graphene[19,35] and n-type P-(NDI2OD-T2) on gold electrodes[36]. It turns out that the different crystalline orientations, namely when the molecules sit on the substrate with the edge (stand-up) or the face (lay-down), play a major role in the final properties of the system. In the case of n-type P-(NDI2OD-T2) thin films it is observed that the injection barrier for electrons is higher in the edge-on than in the face-on devices. In our case, when one-atom thick graphene is used as the electrode, charges are likely to be injected perpendicularly with respects to the molecular layer in contrast with the 30 nm thick gold contacts where the transport of the charge carriers takes place mostly in the parallel direction, resulting in lower contact resistances, as schematically depicted in Figure 2 (g) and (h) respectively. It is important to underline that our result is clearly different from the case of pentacene OFET with graphene electrodes[19], where the beneficial orientation of the interfacial dipole layer is responsible for improved performances. Moreover, it must be stressed also that the pentacene films have the same polarity (p-type) of the graphene electrodes. Nevertheless, the high resistive but still-ohmic region at the electrodes can be considered as a healing factor that hinders the creation of an intense electric field (of the order of MV/cm in ~100 nm long channels for $V_{ds}$=+20 V), limiting the space charge transport in the bulk of the organic channel.

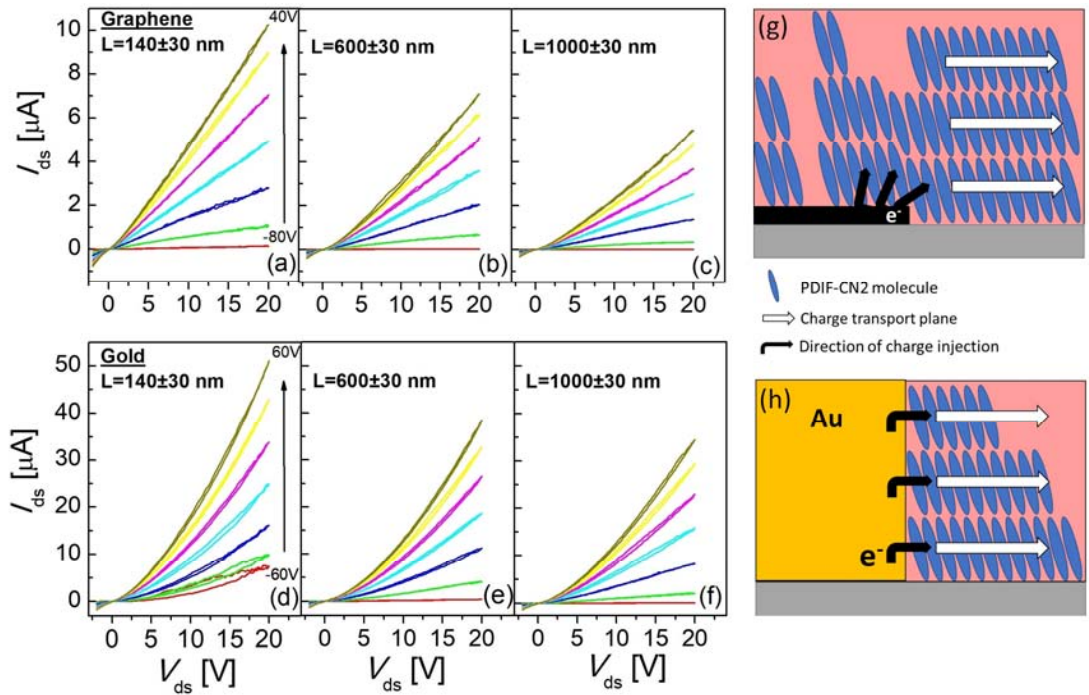

**Figure 2** Output curves obtained in vacuum for different channel lengths $L$ for (a), (b), (c) the graphene-based devices (-80 V≤$V_{gs}$≤+40 V with +20V steps) and (c), (d) (f) the gold-based devices (-60 V≤$V_{gs}$≤+60 V with +20V steps). (g) and (h) Schematic depiction of charge injection for graphene and gold electrodes, respectively

To further check this hypotesis, in Figure 3(a) we plot the output curves in ln-ln scale in order to investigate in deeper details the power-law characteristics $I_{ds} \approx V_{ds}^n$ describing the charge transport in both the architectures. For graphene-based transistors the results show that, even in the case of our shortest channel ($L$=140nm), $I_{ds}$ has a quasi-linear dependence on the entire drain-source voltage interval with $n \approx 1.2$. On the other hand, the power-law for the gold-based devices with comparable channel length appears to be characterized by two distinct regimes depending on the applied drain-source bias (that is equivalent to say the magnitude of the longitudinal electric field) with a slope value $n \approx 1.5$ diverging from the linearity for $V_{ds}$>2 V. This is a typical signature from a SCLC contribution that becomes more evident for decreasing $V_{gs}$ (Figure 3 (b)). The exponent $n$ for $V_{ds}$ > 2 V approaches values exceeding 2 for gate-source bias near -60 V where punch-through currents appear to dominate the charge carrier transport, as we will further discuss below. For $V_{ds}$< 2 V, gold-based devices show slope values similar to those obtained for graphene-based transistors, with $n$ asymptotically approaching 1.2 for increasing $V_{gs}$ biases.

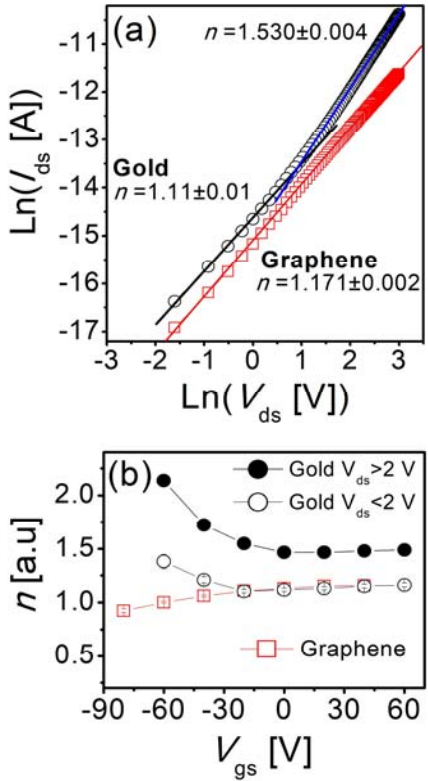

**Figure 3**: (a) Ln-Ln plot of the output curves for $V_{gs}$ = +20 V of the graphene-based device (red squares) and for the gold-based transistor (black circles) with $L$=140±30nm. The $n$ values indicate the slope of the ln-ln curves extracted from linear best fit. (b) Slopes $n$ extracted as a function of the applied gate-source bias in both the architectures for $L$=140±30nm. For gold electrodes devices, the responses for $V_{ds}$ < 2 V (empty circles) and $V_{ds}$ > 2 V (filled circles) are reported.

In Figure 4(a) and (b) we report the transfer curves ($I_{ds}$-$V_{gs}$) for both architectures. As a common feature, highly negative onset voltages ($V_{on}$) are observed with values of about -60 V in the case of gold electrodes and -80 V for graphene electrodes. Large negative threshold values are commonly observed for perylene diimides and especially for PDIF-CN2 when deposited on "bare" (i.e. not passivated) $SiO_2$ substrates, where the presence of charge traps also induces hysteresis in the current characteristics and affects negatively the overall morphology of the organic thin film. These issues are usually overcome by the HMDS functionalization[29]. The effectiveness of our functionalization procedure is confirmed by the ordered polycrystalline morphology of the organic thin film and the absence of hysteresis phenomena in the electrical characteristics. Therefore, we ascribed the highly negative on-set voltages to a partially un-passivated surface, where, likely, a fixed parasitic surface charge density at the $SiO_2$/HMDS interface is still present[37]. Importantly, the negative threshold values do not influence the direct comparison between the graphene and gold–based devices. The advantages in terms of the overall performances in short-channel transistors based on graphene electrodes can be observed comparing the transfer curves reported in Figure 4 (a) and (b). In particular, punch-through currents due to drain-induced barrier lowering are evidently suppressed in graphene-based devices. Indeed, in the case of gold-electrode devices, approaching $V_{gs}$ values near to the onset voltages ($V_{on}$), off-state currents ($I_{off}$) for a fixed $V_{ds}$ bias deteriorate progressively for decreasing $L$, with increasing $I_{off}$ values from $\approx 10^{-9}$A for $L$=140nm to $\approx 10^{-6}$ A for channel lengths approaching 1 μm (Figure 4 (a)). Conversely, off-state currents in graphene based devices appear to sustain the shrinking of the channel length down to $L$ = 140 nm with values fixed in the nA scale (Figure 4 (b)). This obviously affects the on/off ratios ($R_{ON/OFF}$) trends plotted in Figure 4(c) as a function of $L$ where $R_{ON/OFF}$ has been evaluated considering similar device polarizations for both architectures, i.e. considering on-state currents for equivalent effective bias $V'_{gs}= V_{gs} - V_{on}$. Graphene-based transistors essentially show constant $R_{ON/OFF}$ values of the order of $10^2$ in contrast with gold-based devices for which $V_{gs}$ current-modulation decreases towards $10^1$ for $L$<400nm (Figure 4(c)). These results can be explained taking into account the gate-tunability of the work function in monolayer graphene. In particular, considering the measured work function of p-doped graphene at zero bias ($W_F \approx$ -4.9 eV), and the LUMO level of PDIF-CN2 (-4.5eV)[38], a theoretical barrier of $\approx$0.4 eV builds up at the hetero interface. This value depends directly on the applied $V_{gs}$ since $W_F$ for graphene decreases (i.e. becomes more negative) for negative gate voltages, as it was shown by means of Kelvin Probe measurements.[25] As a result, an increasing barrier is expected for decreasing gate-source voltages towards negative onset threshold values, counterbalancing in such a way the drain-induced barrier lowering and suppressing the punch through currents otherwise observed in gold-based transistors. This is further supported by the results reported in Figure 4(d) where $R_{ON/OFF}$ as a function of the applied $V_{ds}$ is plotted for the shortest channel devices ($L$=140nm) and for comparable $V'_{gs}$=100V. Graphene-based transistors takes advantage of the aforementioned barrier modulation with an increasing $R_{ON/OFF}$ for increasing $V_{ds}$, as a consequence of the steadily low $I_{off}$ values while, on the other hand, drain-induced barrier lowering degrades $R_{ON/OFF}$ for increasing longitudinal electric fields in gold-based transistors (Figure 4 (d)). For the sake of completeness, the full set of transfer curves for different $V_{ds}$ are reported in Figure S5 of the supplementary material section.

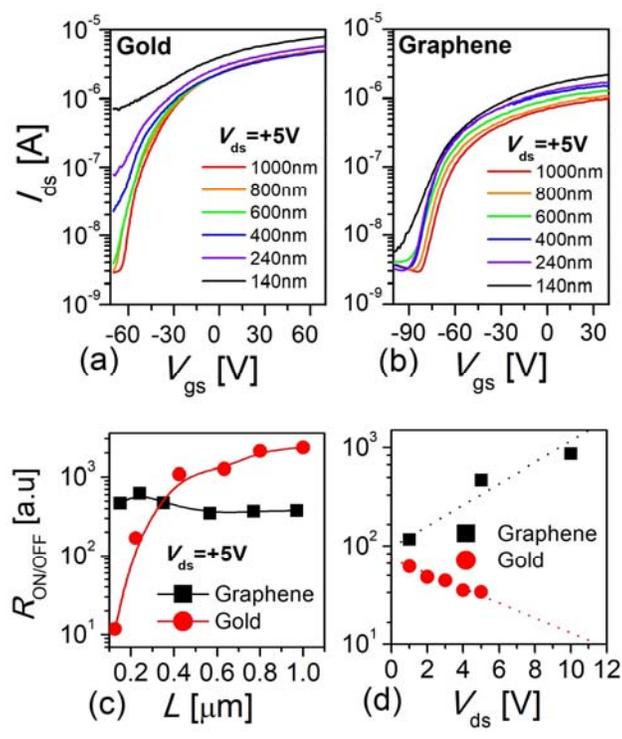

**Figure 4** Transfer curves acquired in vacuum for a fixed drain source bias ($V_{ds}$=+5 V), at different channel lengths for (a) gold and (b) graphene based devices. (c) Comparison of On/Off ratios for the two architectures as function of the channel length $L$ and (d) as function of the applied drain-source voltage $V_{ds}$ for $L$=140nm. In (c) and (d) equal $V'_{gs}= V_{gs}-V_{on}$=100 V have been considered.

In conclusion, p-doped CVD-graphene has been employed as electrode in bottom-contact/bottom-gate n-type OFETs with channel length at the sub-micrometric scale. Experimental results suggest that graphene suppresses short-channel effects thanks to its morphological and electronic properties. A minor contribution of SCLC on the overall response and remarkable improvements in terms of off–state currents are observed when compared to transistors with gold electrodes and equivalent channel lengths. The use of CVD-graphene as electrode is a valuable choice for the development of highly dense support circuitry in all-organic electronic devices, with possible applications for active-matrix-driven Organic Light Emitting Diode (OLED) panels or Organic Light Emitting Transistors (OLET) arrays where mechanical flexibility and low optical absorption are mandatory

**Supplementary Material**

See supplementary materials for further details on the estimation of work function of graphene and electric characterization of the nano-devices.

**Acknowledgments**

The authors want to thank Dr. Mario Barra and Dr. Fabio Chiarella for the helpful discussions on the experimental results.
Dr. Claudia Menozzi is gratefully acknowledged for the support during the fabrication of the devices.
This work was partially supported by European Community through the FET-Proactive Project MoQuaS by contract no. 610449 and by the Italian Ministry for Research (MIUR) through the Futuro In Ricerca (FIR) grant RBFR13YKWX.
Funding from the European Union's Horizon 2020 research and innovation programme under grant agreement No. 696656 – GrapheneCore1 is acknowledged.